\documentclass[]{spie}  

\usepackage{amsmath,amsfonts,amssymb}
\usepackage{graphicx}
\usepackage[colorlinks=true, allcolors=blue]{hyperref}
\usepackage{wrapfig}
\usepackage{lscape}
\usepackage{rotating}
\usepackage{epstopdf}
\usepackage{subcaption}
\usepackage{setspace}
\usepackage{lineno}
\usepackage{multirow}
\usepackage{float}
\usepackage[most]{tcolorbox}

\newcommand{\mum}{$\mathrm{\mu m}$ }
\newcommand{\mltoto}{{\ttfamily MLTOTO} }
\newcommand{\mlsim}{{\ttfamily MLSim} }

\title{Testing of machine learning wavefront sensing algorithms on the Tiny Observatory for Telescope Optimization (TOTO) testbed}

\author[a]{Sanchit Sabhlok}
\author[a,b]{Solvay A. Blomquist}
\author[a]{Maggie Y. Kautz}
\author[a]{Debstuti Biswas}
\author[a,b]{Kevin Derby}
\author[c]{Jaren N. Ashcraft}
\author[a,b]{Hyukmo Kang}
\author[a]{Simran Agarwal}
\author[a,b]{Alexandra Kupersmith}
\author[a]{Adam Schilperoort}
\author[a]{Stephanie Rinaldi}
\author[a]{Kelsey L. Miller}
\author[a]{Kyle J. Van Gorkom}
\author[a]{Corey Fucetola}
\author[a]{Patrick Ingraham}
\author[a]{Ewan S. Douglas}
\author[a,b]{Heejoo Choi}
\author[a,b]{Daewook Kim}
\affil[a]{Steward Observatory - The University of Arizona \linebreak 933 N Cherry Ave, Tucson, AZ 85721 \linebreak}
\affil[b]{Wyant College of Optical Sciences - The University of Arizona \linebreak 1630 E University Blvd, Tucson, AZ 85721}
\affil[c]{Department of Physics, University of California, Santa Barbara, Broida Hall, Santa Barbara, CA 93106}

\authorinfo{Correspondence E-mail: ssabhlok@arizona.edu}

\begin{document} 
\maketitle

\begin{abstract}
Phase retrieval techniques are utilized to correct low order wavefront aberrations originating from misalignments of the optical system in space based telescope concepts. Traditional phase retrieval involves observation of the Point Spread Function (PSF) and a diversity measurement, usually focus diversity although other measures are possible, to reconstruct the incident wavefront at the science detector. We consider a Machine Learning model trained originally on simulated data, and then augmented with real focus diversity data from the Tiny Observatory for Telescope Optimization (TOTO) testbed at the University of Arizona. We then compare the wavefront sensing performance of the Machine Learning model with known truth values of the generated dataset. The model predictions for low order Zernikes on TOTO data after training and validation show a reasonable agreement with the true Zernike coefficients.
\end{abstract}


\section{Introduction}

Alignment of optical systems benefit from the knowledge of the wavefront as it propagates through the system, however it is often not possible to directly measure the wavefront itself. The observable is usually the Point Spread Function (PSF) for the system. This problem is particularly important for systems with Adaptive Optics (AO) which use a Deformable Mirror (DM) to manipulate and modulate the incident wavefront, as the algorithm needs to estimate the wavefront for correction. In the absence of a dedicated wavefront sensor to measure the wavefront, phase retrieval techniques allow recovery of the incident wavefront from measurements of the PSF \cite{Fienup1982:PRAlgorithms}. Specifically, Focus Diversity Phase Retrieval (FDPR) involves observing the PSF at multiple defocus positions to then solve for the wavefront using iterative \cite{GerbhbergSaxton:PR} or non linear optimization \cite{Jurling2014:AlgoDiff} methods.

Phase retrieval is an important problem for large ground based telescopes utilizing AO. It has been used for estimation of Non-common Path aberrations \cite{Ragland2022:PhaseRetrieval}, and PSF reconstruction on sky data for an independent estimate of the PSF in lieu of fitting the PSF to the science image \cite{Ragland2018:PSFR, BeltramoMartin2020:SpherePSFR, Sabhlok2025:PSFR}. Phase retrieval is also utilized for phasing of telescope segments for the James Webb Space Telescope \cite{Acton2022:JWSTPR} and will be utilized on the upcoming Roman Telescope for the Wide Field Imager \cite{Valencia2025:RomanWFIPR} and the Coronagraph Instrument \cite{Marx2025:RomanCGIPR}. 

Current phase retrieval methods are limited by the available computational power for a non-linear optimization method. Another limitation is the dynamic range of the algorithm, which is the maximum amount of deviation in the wavefront before algorithms start running into problems often due to phase wrapping. Machine Learning algorithms rely on training a deep Convolutional Neural Network (CNN) using a training dataset with known truth values for the wavefront that can help increase the dynamic range of current methods. Once trained, a model can also rapidly predict the incident wavefront, which reduces the computational power and time required for a solution. However CNNs do suffer from a "black box" problem, so integration with real hardware needs to be careful and requires support from existing methods. Nonetheless, this approach has shown promising results for computational imaging \cite{Barbastathis2019:DLforCI} and low order wavefront corrections \cite{Allan2020:LOWFSDNN}. The machine learning models also work well in conjunction with existing methods \cite{Kang2020:PhENN} and have been shown to enhance the dynamic range of existing algorithms \cite{Allan2020:ZernikeDNN} in simulations.

In this work, we design and implement a deep CNN model to learn and predict Zernike coefficients of the incident wavefront on the Tiny Observatory for Telescope Optimization (TOTO) testbed at the University of Arizona. We train an initial model using simulated data, and then use transfer learning to train a new model on the testbed data. This is used to predict Zernike coefficients for the incident wavefront on TOTO using defocused PSFs. 

\section{Testbed Design} 

\begin{figure}
	\centering
	\includegraphics[width=\linewidth]{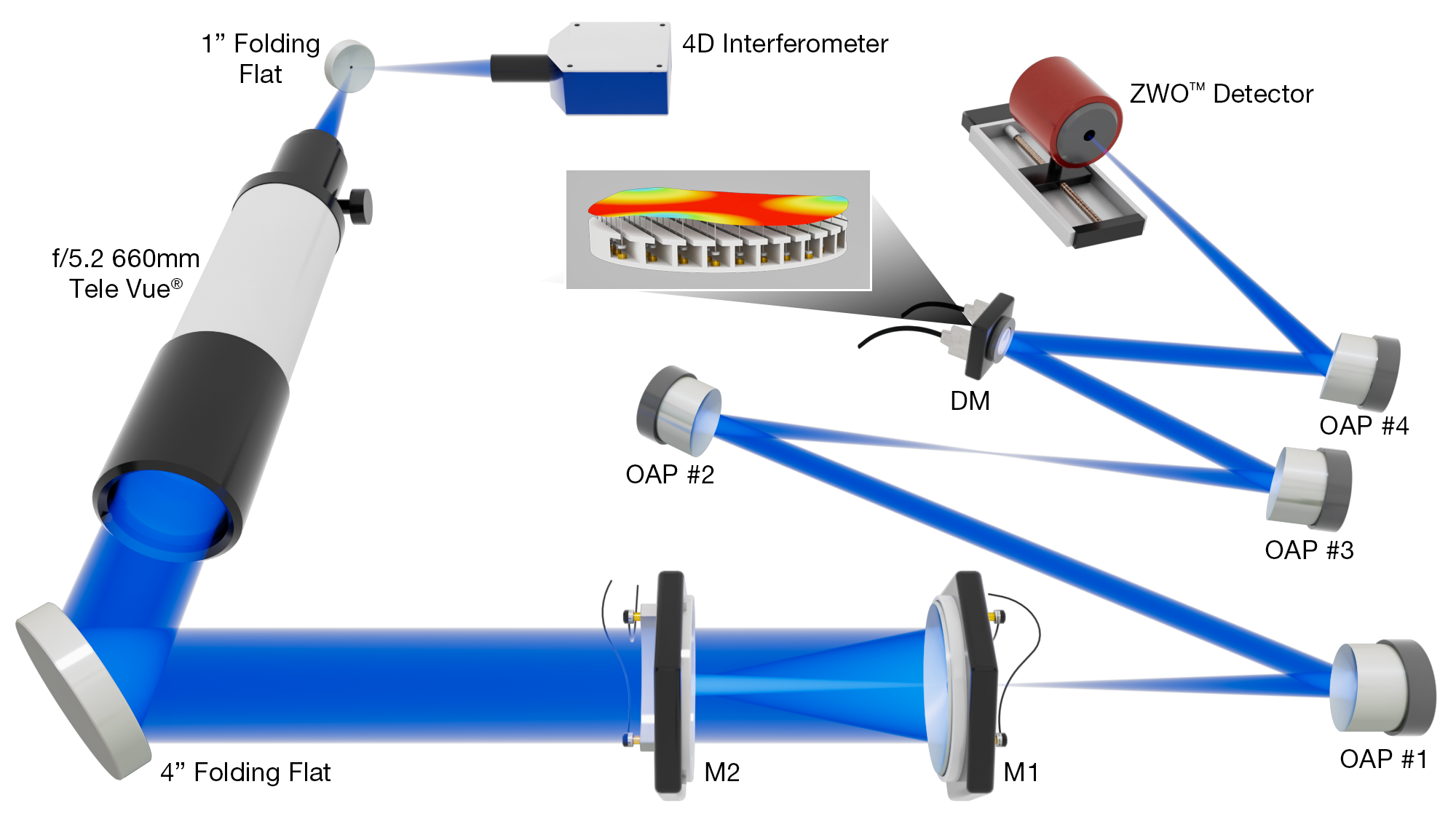}
	\caption{A schematic showing the components on TOTO. Graphic created by Steven Burrows. M1 and M2 are mounted on motorized stages that can control their translation and tip/tilt. The ZWO detector is mounted on a motorized zaber stage aligned with the optical axis to defocus the detector plane.}
	\label{fig:TOTO_Schematic}
\end{figure}

The original TOTO design is discussed in Kang et al. 2025 \cite{Kang2025:TeleSimDesign}. Further details can be found in Blomquist et al. 2026 \cite{Blomquist2026:TOTO}. The current design schematic is shown in Figure \ref{fig:TOTO_Schematic}. The primary light source is the 4D PhaseCam 6000 which operates a 632 nm laser. The beam from the PhaseCam is routed via a 1" folding flat to a TeleVue telescope, which acts as a beam expander and collimator to propagate a 4 inch beam. This beam then goes through a folding flat onto a Ritchey-Chretien telescope \cite{Ashcraft2021:CubeSat}. The secondary mirror on the telescope is mounted on a motorized Zaber stage, which controls M2 translation along X/Y/Z axes, whereas M1 translation is controlled along X/Y/Z axes with a mechanical mount using a screw gauge. The M2 tip-tilt is controlled via a piezoelectric inertia actuators. The deformable mirror (DM) is placed at a pupil conjugate to the primary mirror. The DM used is the ALPAO 97-15 DM which consists of 97 actuators on an 11x11 grid. The DM is used for fine alignment of TOTO and to generate low order modes to be estimated using Phase Retrieval and the deep CNN model. The detector used is the ZWO ASI 6200MM Pro, which uses a Sony IMX 455 sensor. The detector is also mounted on a motorized Zaber stage such that the translational axis of the stage is aligned with the optical axis of the testbed. 

\begin{figure}
	\centering
	\includegraphics[width=\linewidth]{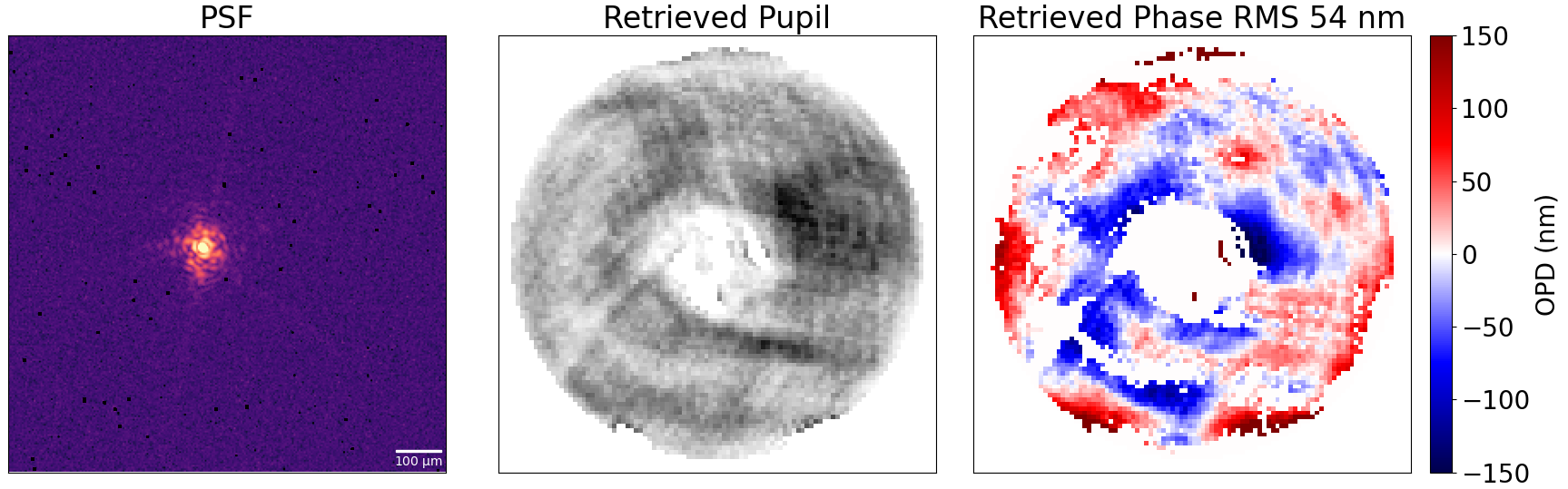}
	\caption{An example of retrieved wavefront amplitude (center) and phase (right) using the FDPR algorithm. An in focus PSF is shown on the left. Note that the in focus PSF is not used for FDPR in our current setup, where only defocused PSFs are used. }
	\label{fig:FDPR2_Example}
\end{figure}

\section{Focus Diversity Phase Retrieval} \label{sec:FDPR}
TOTO has two different methods that can be used to run Focus Diversity Phase Retrieval (FDPR). The detector stage motion can be used to defocus the detector and obtain PSF snapshots. Alternatively, a defocus (i.e., power term) surface shape change can be set at the deformable mirror to get defocused PSF images without moving the detector, which can then be used for Phase retrieval. In this work, we have taken the second approach.  

Focus Diversity is utilized for fine alignment of the testbed and to create an initial baseline for Phase Retrieval performance on the testbed to compare to the Machine Learning algorithms. The FDPR algorithm has been described in detail previously \cite{vanGorkom2021:DMCharacterization, vanGorkom2022:Scoob2}, but we summarize the method here, focusing on aspects specifically relevant for this work.

The algorithm uses the DM to set a defocused Zernike polynomial, normalized to an RMS value of 1 $\mathrm{\mu m}$, and observe the resultant PSF image. These images are used to construct a forward and reverse gradient model of the PSF \cite{Jurling2014:AlgoDiff}, which are simultaneously minimized to obtain the best incident wavefront phase and amplitude. An example of such a phase retrieval model is displayed in Figure \ref{fig:FDPR2_Example}.

We can use FDPR to close a loop to obtain the closed loop PSF. Once the loop is closed, the wavefront is considered to be flat. We can then use the DM to create an incident wavefront which can be decomposed into RMS normalized Zernikes. We use this approach to create wavefronts with known Zernike coefficients, that the MLPR model can train on initially, and then predict the coefficients for validation of model performance as discussed in Section \ref{subsec:MLPRTransferLearning}.

\section{Machine Learning wavefront sensing model using Focus Diversity}
Use of Machine Learning for phase retrieval is motivated by the current state of the art algorithms used for phase retrieval which rely on a technique known as algorithmic differentiation \cite{Jurling2014:AlgoDiff}. These algorithms typically construct a forward model to propagate the wavefront from the pupil to the detector plane, and a reverse gradient model that propagates backwards from the detector to the pupil plane. We can then solve for the wavefront by simultaneously minimizing over the forward and reverse gradient models. The reason why this approach suggests using a machine learning model is because the mathematical basis for the forward and reverse gradient model is identical to that of hidden layers and backpropagation in neural networks.

\begin{figure}
	\centering
	\includegraphics[width=\linewidth]{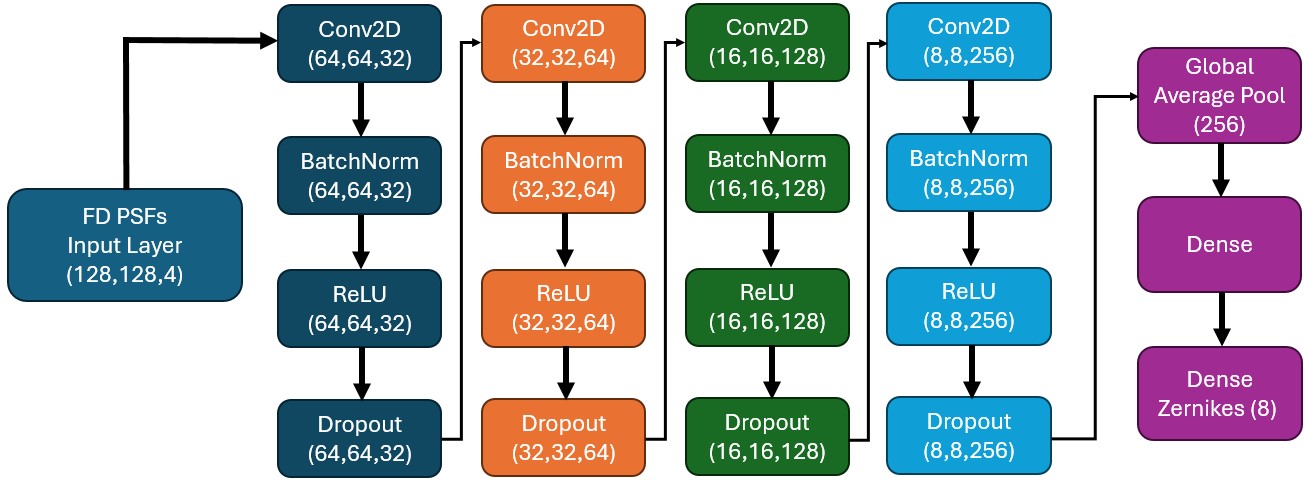}
	\caption{The architecture used for the machine learning model consists of 4 blocks of {\ttfamily Conv2D}, {\ttfamily BatchNormalization}, {\ttfamily ReLU} and {\ttfamily Dropout} with decreasing kernel sizes. This is followed by a {\ttfamily Global2DAveragePool} and a {\ttfamily Dense} layer which connects to the output Zernike vector. The number of Zernikes in the final layer depends on the training data. }
	\label{fig:tensorflow_model_architecture}
\end{figure}

\begin{figure}[b]
	\centering
	\includegraphics[width=\linewidth]{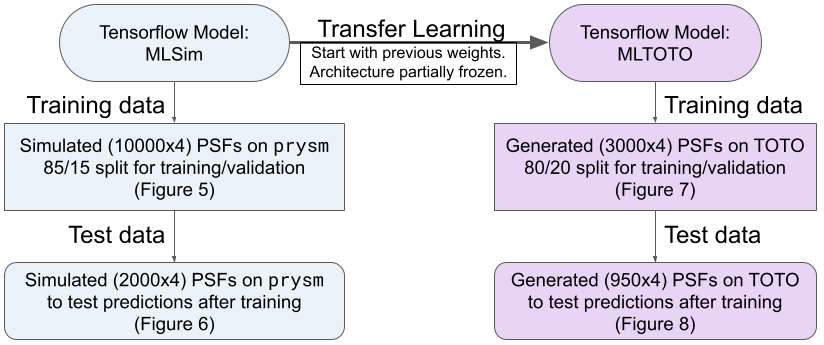}
	\caption{Schematic describing the training of the MLPR models. Both models the same architecture as shown in Figure \ref{fig:tensorflow_model_architecture}. We first train \mlsim on simulated data (Figure \ref{fig:model_sims_training_plots}) and predict Zernikes on a previously unseen simulated dataset (Figure \ref{fig:model_predictions_sims}). The weights from \mlsim are then transferred to \mltoto, which is retrained on dataset taken on TOTO (Figure \ref{fig:model_toto_training_plots}) and then predicts Zernikes for a previously unseen dataset taken on TOTO (Figure \ref{fig:model_predictions_TOTO}).  }
	\label{fig:MLPR_schematic}
\end{figure}

We build a neural network architecture using {\ttfamily tensorflow} python package \cite{tensorflow2015-whitepaper}. We train the model in two stages as shown in Figure \ref{fig:MLPR_schematic}. First we train the model on simulated data generated using the {\ttfamily prysm} physical optics propagation library \cite{prysm:JOSS, Prysm2}, which we call the \mlsim model. Then we use the trained model to transfer learning using focus diversity dataset from TOTO, which we call the \mltoto model. All dataset generation and ML model fitting here was carried out on a laptop with an Nvidia GeForce RTX 4060 Max-Q / Mobile graphic card and 32 GB of RAM. The training of the simulated dataset and the lab dataset both take $\sim$ 2 seconds per Epoch, for a total of $\sim$ 16 minutes for 500 epochs. We now discuss the training of the model using simulated and real dataset and its application on TOTO.

\subsection{Training \mlsim with the Simulated Dataset}
The input layer comprises of a set of PSFs with a given set of defocus values. As discussed previously in Section \ref{sec:FDPR}, the PSF is defocused by setting a Zernike polynomial command at the DM, where the Zernike basis is normalized to RMS of 1 \mum. The simulated dataset aims to replicate this setup. We use {\ttfamily prysm} to generate arbitrary wavefronts using a set of low order Zernikes Z4-Z11, all normalized to RMS of 1 \mum. We use the Noll indexing for the Zernikes so Z4 corresponds to (2,0) (defocus) and Z11 to the (4,0) (primary spherical) term. The range of coefficients used in the simulation is $\pm$ 50 nm, or 0.05 \mum. This is a small range of Zernikes to train over and predict, however future work will explore the performance of an ML model over a larger dynamic range.

The optical system used to generate the training dataset uses the following parameters. The monochromatic PSFs are generated at a wavelength of 632.8 nm, which matches the wavelength of the laser light source on TOTO. \mlsim also uses the appropriate pupil size and f-number to match the TOTO design. The values of defocus coefficients chosen for this experiment are (-0.5, -0.2, 0.2, 0.5) \mum. The dataset is normalized for all PSFs, such that the peak pixel value for all PSFs is 1. We then add a gaussian noise term with a centroid of 1e-3 and standard deviation of 1e-4 to this dataset, to approximate detector noise on the real dataset. This forms the input layer for the deep CNN model, which is concatenated to a shape of (N$_{\mathrm{PSFs}}$, H$_{\mathrm{image}}$, W$_{\mathrm{image}}$, Defocus), which is a {\ttfamily numpy.ndarray} of shape (10000, 128, 128, 4). 

The training dataset is passed to \mlsim for fitting where it undergoes a split to divide into a training and validation dataset. We have chosen a fraction of 85\% for \mlsim, dividing the initial dataset randomly into 85\% training and 15\% validation data. \mlsim is then trained on the training dataset, and tested on the validation dataset. For each epoch, \mlsim evaluates the loss function and the correlation coefficient for both training and validation datasets. \mlsim trains to minimize the \textit{validation loss function}, which here is chosen to be the {\ttfamily tf.keras.losses.MeanSquaredError} function. Mean Squared Error (MSE) has been chosen here over other loss functions such as the Huber or cross entropy loss. This is motivated by the use of the same error function when performing phase retrieval using algorithmic differentiation. We use an {\ttfamily Adam} optimizer to train \mlsim, starting with a learning rate of 10$^{-5}$ and reducing the learning rate if the validation loss plateaus over 50 epochs of training. \mlsim trains on the simulated dataset for 500 epochs. 

Both \mlsim and \mltoto have the same underlying architecture, shown in Figure \ref{fig:tensorflow_model_architecture}, which we refer to as the deep CNN architecture. The network comprises of a feed forward model with no skip connections. The first four layers of the network apply the same operations - 
\begin{enumerate}
    \item {\ttfamily Conv2D} - This is a 2D convolution layer which acts on the input from the previous layer. Across the 4 layers, the Conv2D starts with a kernel of (64,64,32) which identifies larger scale features and then proceeds to smaller kernel sizes of (32, 32, 64), (16, 16, 128) and finally (8, 8, 256) to identify smaller features. The Conv2D layer also uses an L1-L2 regularizer to avoid overfitting minor features. The anomalies due to minor features are better captured by the {\ttfamily GlobalAveragePool2D} layer.
    \item {\ttfamily BatchNorm} - Applied batch normalization to results of the Conv2D convolution and feeds forward to the activation layer.
    \item {\ttfamily ReLU} - We utilize a Rectified Linear Unit (ReLU) activation layer. Use of this activation function allows for healthy gradient flow throughout the architecture as a sigmoid activation could result in vanishing gradients. We also find this performs better than a sigmoid activation combined with a skip layer. 
    \item {\ttfamily Dropout} - We add a dropout layer to the model to avoid overfitting through the convolution layers. We use a dropout value of 0.3 for the simulated and transfer learning datasets. 
\end{enumerate}

\begin{figure}
	\centering
	\includegraphics[width=\linewidth]{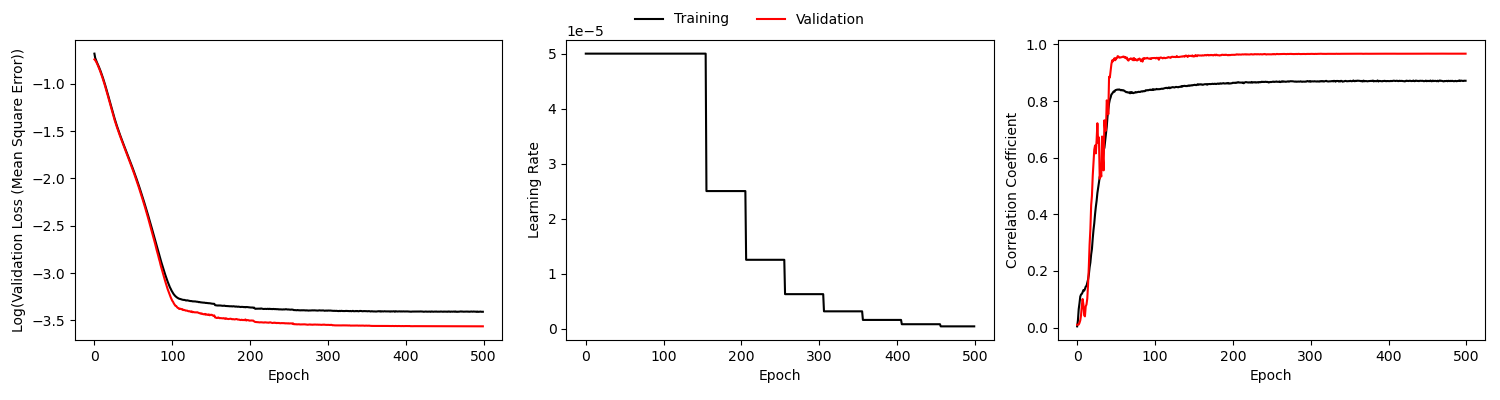}
	\caption{Training parameters for the simulated data. The model is trained by minimizing the Validation loss, which is a mean squared error function (left). The model also uses a callback method to reduce by half the learning rate if the loss function plateaus (middle). We also track the  Pearson correlation coefficient between truth and validation Zernike coefficients over the course of training (right). }
	\label{fig:model_sims_training_plots}
\end{figure}

The model architecture then passes forward through a {\ttfamily GlobalAveragePool2D} layer, which averages features over a downsampled version of the feature layers. The goal here is to average over features to detect low order Zernike features robustly and prevent overfitting. When training on data from the testbed, the layer provides an additional layer of robustness as it suppresses the impact of jitter on individual PSFs. 

This layer is then connected to a {\ttfamily Dense} layer, which connects to a layer of Zernikes, which is also our output. The size of the final layer is controlled by the number of Zernikes being predicted in the model. Note that it is possible to train a model using simulated data which utilizes a higher number of Zernikes than it predicts. In this work however, the simulated training dataset and the predicted Zernike coefficient vector have the same number of terms. 

\begin{figure}
	\centering
	\includegraphics[width=\linewidth]{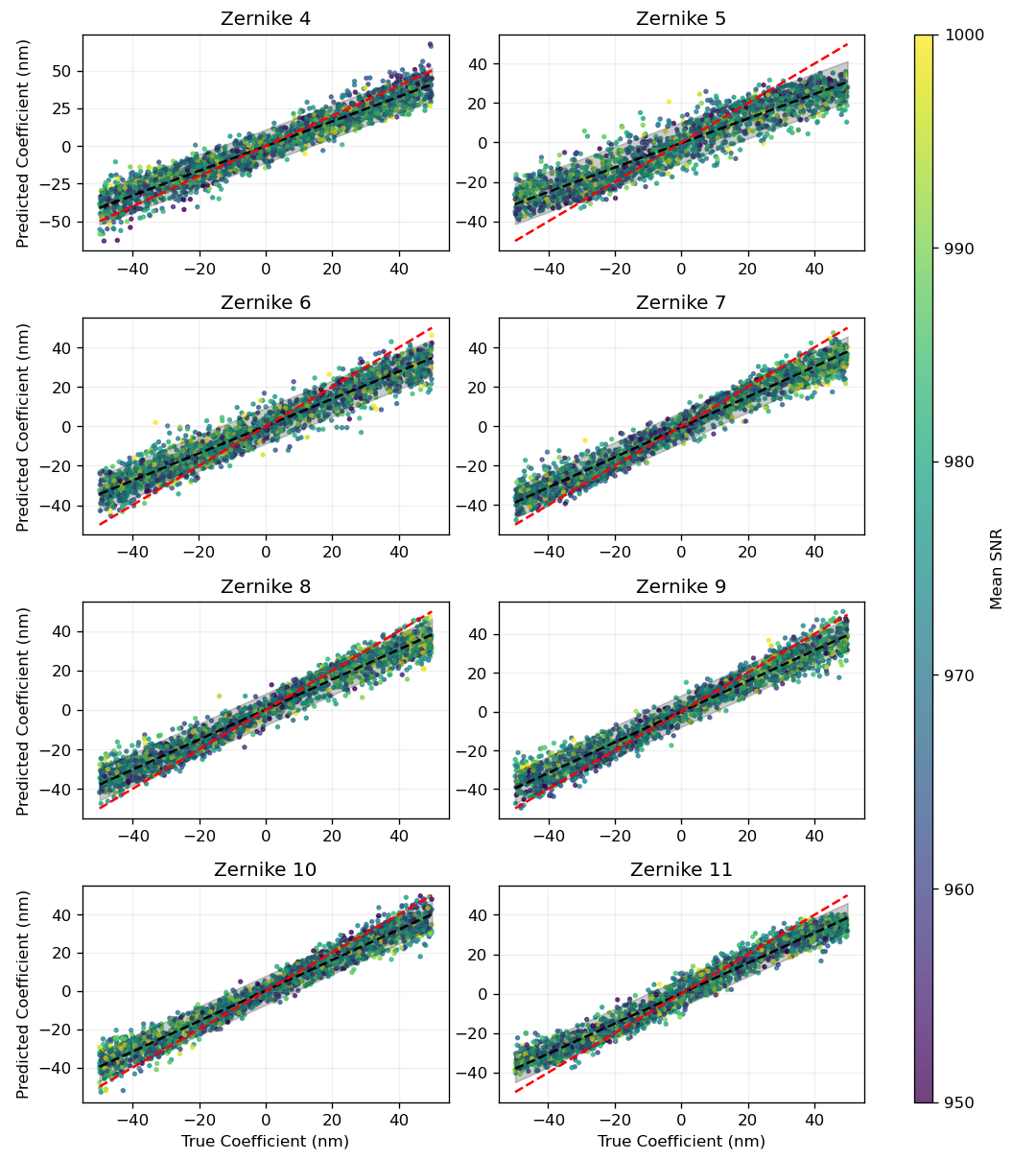}
	\caption{Predictions from model trained on simulated data on a set of 2000 simulated PSFs previously not seen by the model during training. Each datapoint is the predicted Zernike coefficient using 4 defocused PSFs vs. the true value of the coefficient. The x$=$y line is shown with a red dashed line whereas the black dashed line shows the best linear fit, and the shaded black region shows the 90th percentile around the linear fit. We find that the magnitude of Z5, Z6 and Z11 is under-predicted at the edge of the dynamic range. }
	\label{fig:model_predictions_sims}
\end{figure}

\subsection{Results from Simulated Dataset Training}
We investigate \mlsim training plots for any irregularities, as shown in Figure \ref{fig:model_sims_training_plots}. We find that the training and validation loss functions all behave smoothly converging to a final value. The same is true for the net correlation coefficient for all predicted Zernikes. The learning rate plot shows a systematic decrease over the course on training, triggered by the plateauing of validation loss, with a patience of 50 epochs. 

After training and validating \mlsim, we generate a new dataset that the model hasn't seen for training or validation before. The new dataset consists of 2000 set of defocused PSFs. \mlsim then predicts the Zernike coefficients for the new dataset, and the predictions are compared to true values. 

The results are shown in Figure \ref{fig:model_predictions_sims}. We plot the true and predicted values for each set of defocused PSFs, color coded by the mean SNR of the 4 PSFs used to predict the Zernike coefficient. We also do a linear fit on the data and plot a band of the 90th percentile of data points shaded in black. We observe a linear trend, and don't find a significant dependence on the SNR for the predicted coefficients. \mlsim systematically underpredicts Z5, Z6 and Z11 at the edge of the dynamic range of $\pm 50$ nm, since the x$=$y line lies beyond the 90th percentile of the linear fit for these coefficients. This is acceptable as the training dataset from TOTO has a smaller range of Zernike coefficients of $\pm 30$ nm, where \mlsim is expected to behave reasonably well. 

The resultant model and logs are saved. These model weights are then used to transfer the learning to \mltoto trained using data from TOTO as described next. 

\subsection{Results from Transfer Learning on data from TOTO} \label{subsec:MLPRTransferLearning}
The TOTO dataset used for transfer learning comprises of 3950 set of defocused PSFs taken on a single day. This is important as the optical alignment of the testbed can vary slightly day to day, so it is important to generate a DM flat at the start of data collection and collect the entire dataset with minimal changes to the testbed. 

We use FDPR as described in Section \ref{sec:FDPR} in closed loop control to flatten the wavefront using the DM. This takes 3-5 iterations but we run closed loop for 10 iterations to get a good flat. We then generate a set of 8 random Zernike coefficients between -30 and 30 nm and generate the wavefront by adding these modes. Then, we set the DM surface to (-0.25, -0.1, 0.1, 0.25) \mum of defocus, which correspond to (-0.5, -0.2, 0.2, 0.5) \mum of defocus, assuming ideal actuators and perfect alignment of the DM. We get a 256x256 image of the PSF for each defocus position at an SNR of $\sim 100$. We then center and crop the image down to 128x128 and normalize it so the peak pixel value is equal to 1. This dataset is then used for transfer learning of \mltoto. 

\begin{figure}
	\centering
	\includegraphics[width=\linewidth]{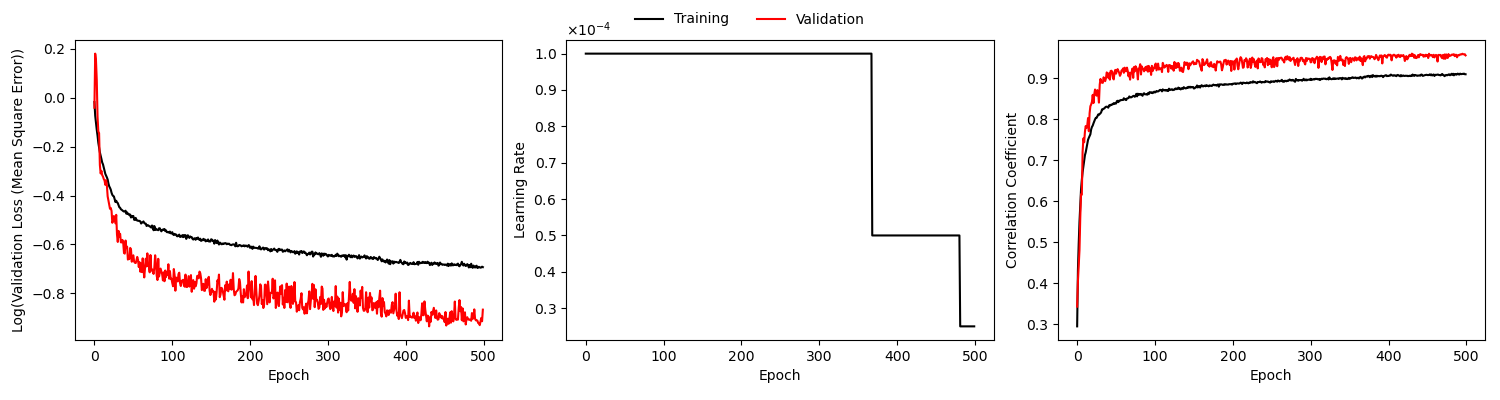}
	\caption{Training parameters for \mltoto trained on testbed data. \mltoto follows the same architecture as \mlsim, but the weights for the first two blocks are frozen during training. \mltoto shows considerably higher variance in validation loss and the Pearson correlation coefficient than the simulated data, although this is likely due to the smaller training dataset of 3000 PSFs. The learning rate takes more epochs to start reducing since the validation loss does not plateau for a long time. This suggests that more data and longer training times can result in a better trained model.  }
	\label{fig:model_toto_training_plots}
\end{figure}

For transfer learning, we start with the weights of \mlsim and an identical architecture, as shown previously in Figure \ref{fig:tensorflow_model_architecture}. We freeze the first two blocks of the architecture, as the larger convolution blocks likely trace larger spatial features in the PSF, and we would like to retain the training from \mlsim. We unfreeze the weights for the remaining layers and train this model using the lab dataset.

\begin{figure}
	\centering
	\includegraphics[width=\linewidth]{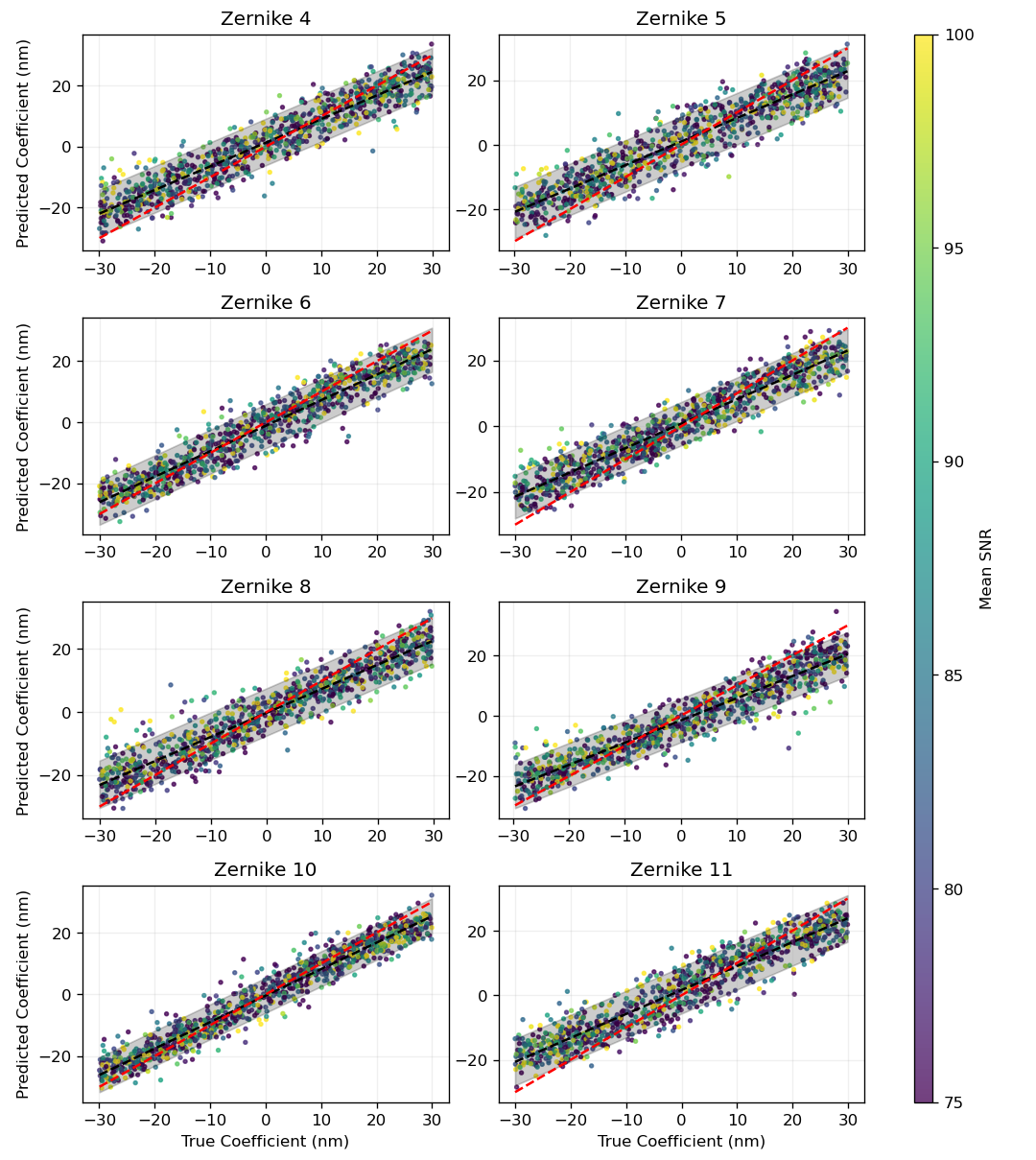}
	\caption{Predictions from \mltoto on a set of 950 simulated PSFs previously not seen by the model during training. The figure markings are identical to that in Figure \ref{fig:model_predictions_sims}. However, the extent of the X axis is reduced to the dynamic range of $\pm 30$ nm, whereas the SNR colorbar now spans 75--100, which is an order of magnitude smaller than Figure \ref{fig:model_predictions_sims}.}
	\label{fig:model_predictions_TOTO}
\end{figure}

The results from the training are shown in Figure \ref{fig:model_toto_training_plots}. Compared to the training for the simulated dataset the validation loss and training loss show significantly higher variance per epoch throughout the training process. The same is true for the correlation coefficient. 

\mltoto uses mostly identical training hyperparameters to that for \mlsim, except for learning rate, for which we start with 1e-4 and reduce if the validation loss plateaus for 50 epochs. The training dataset uses 3000 out of the 3950 PSFs collected for the experiment, splitting it into training and validation datasets of 2400 and 600 set of defocused PSFs respectively. The remaining set of 950 PSFs are used to predict the Zernike coefficients after training. The results of these predictions are shown in Figure \ref{fig:model_predictions_TOTO}. 

\mltoto shows a linear correlation for all individual Zernikes. Unlike the simulated data, the x$=$y line is not significantly offset from the linear fit and the 90th percentile band at the edge of the dynamic range, which is likely helped due to the smaller dynamic range of the training dataset when compared with the simulated dataset. Looking at the SNR distribution for the predictions, we do see here that most outliers outside the 90th percentile tend to have lower mean SNR for the defocus PSFs, although this trend likely needs more data to confirm. The exceptions where high SNR points exceed the 90th percentile lie closer to the edge of the dynamic range than the center. More importantly, we note that while the simulated dataset was trained with a mean SNR closer to 1000, the lab dataset only required SNR $\sim$ 100 to allow sufficient transfer learning. This is promising as simulated datasets can be generated with arbitrarily high SNR, but real data can have lower SNR values for adequate performance. 

\section{Summary and Future Work}

We have trained a Convolutional Neural Network on a simulated dataset and transferred the learning from the model to train a new model on data taken from the TOTO testbed. Both models were trained to predict 8 lower order Zernikes from Z4 through Z11, and then tested on a simuilated and testbed datasets. Both model predictions show a linear trend compared to the true values. The truth values are within the 90th percentile from the linear fit for all 8 Zernikes. Future work needs to improve the bounds on the error of the estimate Zernike coefficients, but the current performance on a testbed is promising.

We expect gains in performance from fine tuning the hyper-parameters for both \mlsim and \mltoto. Having verified that \mltoto can predict Zernikes on lab data, the next steps are to use the best performing model to run a low order closed loop on TOTO. Once this is functional, we can generate further datasets to investigate the performance of the model on the dynamic range and SNR of the training and simulated dataset and the model architecture. 

\section*{ACKNOWLEDGMENTS}

Portions of this research were supported by funding from the Technology Research Initiative Fund (TRIF) of the Arizona Board of Regents and by generous philanthropic donations to the Steward Observatory of the College of Science at the University of Arizona. J.N.A was supported by NASA through the NASA Hubble Fellowship grant \#HST-HF2-51547.001-A awarded by the Space Telescope Science Institute, which is operated by the Association of Universities for Research in Astronomy.

\bibliography{spie} 
\bibliographystyle{spiebib} 

\end{document}